# Anomalous magnetic moment and Compton wavelength

Raji Heyrovska

J. Heyrovský Institute of Physical Chemistry, Academy of Sciences of
the Czech Republic, Dolejškova 3, 182 23 Prague 8, Czech Republic.

Raji.Heyrovska@jh-inst.cas.cz

**Abstract**. The relativistic and quantum theoretical explanations of the
magnetic moment anomaly of the electron (or proton) show that it is a
complicated function of the fine structure constant. In this work, a
simple non-relativistic approach shows that the translational motion of
the particle during its spin is responsible for the observed effects.

**Key words**: anomalous magnetic moment, spin, fine structure
constant, translational motion during spin, electron radius

The magnetic moment anomaly ($a_e$) of an electron (or $a_p$ for a proton)
concerns the observed higher value ($\mu_e$) for a free electron (or $\mu_p$ for a
proton) than the expected value equal to the Bohr magneton ($\mu_B$) (or
$\mu_N$, the nuclear magneton for the proton). An introduction to the topic
and the bibliography can be found in [1]. The relativistic and quantum
theoretical explanations given suggest that the anomaly is a function
of the fine structure constant. In this work, a simple non-relativistic
approach shows that the translational motion of the particle during its
spin is responsible for the higher magnetic momenta of the particles.

For the case of an electron (subscript e), (note: all equations hold for a proton with subscript p in place of e)

$$\mu_e = (g_e/2)\mu_B = (1 + a_e)\mu_B > \mu_B \qquad (1)$$

where $(g_e/2)$ ( >1) is the anomalous g-factor for the electron and $\mu_B$, the Bohr magneton is given by,

$$\mu_B = (\hbar e/2m_e) \qquad (2a)$$

$$[= (e^2/\kappa v_\omega)(e/2m_e) = (ec/2\pi)(\lambda_{C,e}/2)] = (ec/2)r_e \qquad (2b)$$

where $\hbar = h/2\pi = e^2/\kappa\alpha c$, $\kappa = 4\pi\varepsilon_o$ is the electrical permittivity of vacuum and $\alpha$ is the fine structure constant. The equivalent terms introduced here in the square brackets consist of $\hbar = e^2/\kappa v_\omega$, the angular momentum of spin and $v_\omega = \alpha c$, the velocity of spin (as a result of the torque due to electric and magnetic fields [2]), the Compton wavelength,

$$\lambda_{C,e} = h/m_e c = d_e = (2\pi e^2/\kappa v_\omega m_e c) = 2\pi r_e \qquad (3)$$

equal to the distance $d_e = 2\pi r_e$, $r_e = e^2/\kappa v_\omega m_e c = \hbar/m_e c$, considered here as the radius of the electron and $I_\omega = \hbar/c = m_e r_e$, the rotational moment of inertia. Thus $d_e$ is the distance a point on the circumference of the electron would cover during one spin of the electron around an axis through its center.

Equation (1) can therefore be written as,

$$\mu_e = \hbar\gamma \ [= (g_e/2)(e^2/\kappa v_\omega)(e/2m_e) \tag{4a}$$

$$= (ec/2\pi)(d_{e,f}/2)] = (g_e/2)(ec/2)r_e] \tag{4b}$$

where $\gamma_e = (g_e/2)(e/2m_e)$ is the gyromagnetic ratio [1] of the electron, and the distance $d_{e,f}$ is given by,

$$d_{e,f} = (g_e/2)d_e = d_e + \delta d_e = (1 + a_e)(2\pi e^2/\kappa v_\omega m_e c) \tag{5}$$

where $\delta d_e$ is the translational displacement of the electron during its spin. Thus, the g-factor and the magnetic moment anomaly stand respectively for,

$$(g_e/2) = (1 + a_e) = d_{e,f}/d_e = (1 + \delta d_e/d_e) \tag{6}$$

$$a_e = \delta d_e/d_e \tag{7}$$

where $d_e$ depends on the fine structure constant, see equation (3).

**Acknowledgement:** This work was financed by Grant

101/02/U111/CZ